\def\etal{et al.~\/}

\def\ie{{\it i.e.~\/}}

\def\lya{Ly$\alpha$}
       
\def\ltsima{$\; \buildrel < \over \sim \;$}
\def\simlt{\lower.5ex\hbox{\ltsima}}
\def\gtsima{$\; \buildrel > \over \sim \;$}
\def\simgt{\lower.5ex\hbox{\gtsima}}
\newcommand{\be}{\begin{equation}}
\newcommand{\ee}{\end{equation}}
\newcommand{\ba}{\begin{eqnarray}}
\newcommand{\ea}{\end{eqnarray}}

\documentstyle[11pt,epsfig,aaspp4]{article}
\tighten
\singlespace

\lefthead{Ferrara, Pettini \& Shchekinov}
\righthead{Metal Mixing}    

\begin{document}

\title{Mixing Metals in the Early Universe}

\author{Andrea Ferrara$^1$, Max Pettini$^{2}$, \& Yuri Shchekinov$^{1,3}$}
\affil{
$^1$Osservatorio Astrofisico di Arcetri, Firenze, Italy \\
$^2$Institute of Astronomy, Madingley Road, Cambridge CB3 0HA, UK\\ 
$^3$Department of Physics, Rostov State University, 
Rostov on Don, Russia}
\begin{abstract}

We investigate the evolution of the metallicity of the
intergalactic medium (IGM) with particular emphasis on its
spatial distribution. We propose that metal enrichment occurs as
a two step process. First, supernova (SN) explosions eject metals
into relatively small regions confined to the surroundings of
star-forming galaxies. From a comprehensive treatment of blowout
we show that SNae by themselves fail by more than one order of
magnitude to distribute the products of stellar nucleosynthesis
over volumes large enough to pollute the whole IGM to the
metallicity levels observed. Thus, a additional (but as yet
unknown) physical mechanism must be invoked to mix the metals on
scales comparable to the mean distance between the galaxies which
are most efficient pollutants. From this simple hypothesis we 
derive a number of testable predictions for the evolution of the
IGM metallicity. Specifically, we find that: {\it (i)} the
fraction of metals ejected over the star formation history of the
universe is about 50\% at $z=0$; that is, approximately half of
the metals today are found in the IGM; {\it (ii)} if the ejected
metals were homogeneously mixed with the baryons in the universe,
the average IGM metallicity would be $\langle Z
\rangle=\Omega_Z^{ej}/\Omega_b \simeq 1/25 Z_\odot$ at $z=3$.
However, due to spatial inhomogeneities,  the mean of the
distribution of metallicities in the diffusive zones has a wide
(more than 2 orders of magnitude) spread around this value; {\it
(iii)} if metals become more uniformly distributed at $z\simlt
1$, as assumed, at $z = 0$ the metallicity of the IGM is narrowly
confined within the range $Z \approx 0.1 \pm 0.03 Z_\odot$.
Finally, we point out that our results can account for the
observed metal content of the intracluster medium.

\end{abstract}


\section{Introduction}

Primordial nucleosynthesis enriched gas in the universe with 
the light elements He, D, and Li. It is only when the first galaxies 
and their stars appeared that heavier elements could be synthesized 
and, in some cases, ejected into the intergalactic
medium. In currently popular models of galaxy formation based on 
hierarchical clustering the first galaxies to form were 
low mass systems with such shallow potential wells that a few 
supernovae could deposit sufficient kinetic energy to expel the entire 
interstellar medium of the galaxy (Ferrara 1998).
In this way, these initial episodes of star formation in the universe
(sometimes referred to as Population III) may have 
enriched the intergalactic medium (IGM) to an average metallicity
$Z_{\rm IGM} \approx 10^{-4} Z_\odot$ (Miralda-Escud\'e \& Rees 1997, 
Nath \& Trentham 1997, Gnedin \& Ostriker 1997, Ciardi \etal 2000b),
comparable to that of the most metal-poor stars in the halo of our Galaxy 
(Ryan, Norris, \& Beers 1996).

Most of the metals in the universe were presumably produced in 
larger collapsed structures at redshifts $z \simlt 10$. In general such 
galaxies were better able to retain the products of stellar nucleosynthesis
and could therefore achieve the abundance levels observed today. 
It is also likely that some metal-enriched gas escaped into the IGM,
but it is far from clear by which process 
and to what degree the metals became 
distributed over large volumes, far from their production sites. 
These are the questions we consider in the present paper.
Recent numerical hydrodynamic simulations of large-scale structure formation 
(Hellsten \etal 1997; Rauch \etal 1997; Zhang \etal 1998) have shown that baryons 
in the universe are distributed in a network
structure (the ``cosmic web'')
with galaxies located at the high overdensity peaks.
This implies that most of the volume is occupied by voids filled with gas 
at, or below, the mean cosmic density; 
such gas can be viewed as the long-searched-for true
intergalactic medium. 
Since the voids are very large, with typical dimensions of several Mpc,
their pollution by heavy elements produced in SN explosions is 
far from a trivial problem. 
In any case there must have been a protracted epoch when the
metal content of the IGM was highly patchy.

Observationally, our best view of the IGM is still provided by the 
\lya\ forest in the spectra of distant QSOs (Sargent et al. 1980)
and the detection of metals in the forest ranks as one of the most 
significant discoveries made possible by the Keck 
telescopes (Cowie \etal 1995; Tytler \etal 1995).  From 
an analysis of these data Hellsten et al. (1997) and
Rauch, Haehnelt, \& Steinmetz (1997) 
concluded that the measured column density ratios $N$(C~IV)/$N$(H~I)  
imply that typically [C/H]~$ \simeq -2.5$ at $z \simeq 3$, 
with a one order of magnitude dispersion 
in the metallicity of different clouds 
about this mean value.\footnote{In the usual notation,
[C/H] = log (C/H) $-$ log (C/H)$_{\odot}$.} 
At lower redshifts ($z=0.3-0.8$) Barlow \& Tytler (1998) using
HST/FOS spectra, with reasonable assumptions concerning the
ionization correction and the line clustering properties, conclude
that metallicities are as high as [C/H]$\simgt -1.3$, roughly an order
of magnitude larger than the value at $z=2.5$.   
These measurements, however, still refer to overdense regions
of the universe, traced by \lya\ clouds with column densities in
excess of log~$N$(H~I) = 14.5\,.
The situation is far less clear-cut when we turn to the 
voids---or log~$N$(H~I)$ < 14.0$---where 
the observations are very challenging 
even with a 10-m telescope. 
Two studies have addressed this problem, with conflicting 
results.
Lu \etal (1998) applied a stacking technique 
to nearly 300 C~IV regions in 
QSO spectra but still found no composite signal.
They interpreted this non-detection as evidence 
for a highly non-uniform degree of metal enrichment at $z \simeq 3$,
with the voids having [C/H]~$\simlt -3.5$\,.
On the other hand,
Cowie \& Songaila (1998) used
a pixel-to-pixel optical depth technique
to conclude that the average 
C~IV/H~I ratio remains essentially constant
over the full range of neutral hydrogen column densities tested, 
down to log~$N$(H~I)~$\simeq 13.5$\,.
If this is indeed the case, the ejection and transport of metals away 
from galaxies must have been much more efficient than envisaged.
Most recently, Ellison \etal (1999, 2000) 
have re-examined the problem and found
that both approaches suffer from limitations which had not been properly 
taken into account in previous analyses. In their view, whether
there are metals in the voids remains an open question.

On the theoretical side too, only a limited amount of work 
has been done on this topic. 
The common assumption that supernova driven winds may 
be able to distribute metals over large distances
has been shown to be too 
simplistic and does not stand up to close quantitative scrutiny
(MacLow \& Ferrara 1999; Murakami \& Babul 1999;
D'Ercole \& Brighenti 1999). 
These studies have come to the conclusion that
efficient blowout is likely to be inhibited by the galaxy ISM and,
at least at high redshift where densities are higher, by the pressure of the 
surrounding intergalactic gas (Babul \& Rees 1992; Ciardi \& Ferrara 1997).
Gnedin \& Ostriker (1998) numerically simulated
the IGM enrichment in a CDM+$\Lambda$ cosmological model. 
They found that 
the mean metallicity of the universe at $z=4$ is about 1/200 $Z_\odot$
but the variations around this value are large and dependent on the 
overdensity.  At this epoch the enrichment was incomplete, with some 
regions of the universe still of pristine composition. Their simulations, 
however,
end at $z=4$ so that a direct connection with the nearby universe
is difficult. In addition, quantitative conclusions on the transport 
mechanism are affected by uncertainties arising from insufficient 
numerical resolution so that, for example, they are unable to
resolve the snowplough phase of 
explosions. 
To alleviate the problem of inefficient blowout
Gnedin (1998) developed the idea
that metals are predominantly transported following merger events, a
mechanism which would also predict a highly inhomogeneous distribution
of heavy elements in the IGM at $z = 4$. However, 
the merger history has not yet
been followed to more recent epochs.

In this paper we make explicit predictions for 
the IGM metallicity evolution by
suggesting a two-step mechanism for the 
transport and mixing of heavy elements.
In our scenario, 
SN explosions first eject metals 
over a relatively small region in the surroundings
of the host galaxy;
subsequently, a yet unknown diffusion process 
transports and mixes these elements over much larger scales. 
The plan of the paper is as follows. In \S2 
we consider the conditions for gas 
ejection from galaxies, paying particular attention to the treatment of 
blowout, which we show 
to be quite a rare occurrence in massive galaxies.
In \S3 we explicitly derive the metallicity evolution of the IGM
for a particular (Cold Dark Matter) cosmological model. 
The results are discussed
in \S4 and a brief summary (\S5) concludes the paper.

\section{Metal Ejection by Galaxies}

It is likely that the first episodes of star formation in the universe
had a dramatic impact on the small mass galaxies in which they occurred.
The combined effects of supernova explosions in relatively small volumes
lead to the formation of superbubbles (SBs) which can result in 
partial (blowout) or even complete (blowaway) removal of the interstellar 
medium (Kovalenko \& Shchekinov 1985; MacLow \& McCray 1988; 
Ciardi \& Ferrara 1997; Ferrara 1998; MacLow \& Ferrara 1999).
The resulting large scale outflows (galactic superwinds) 
have been observed directly
in local starbursts (Gonz\'{a}lez Delgado et al. 1998; Heckman et 
al. 1998) and in Lyman break galaxies
at $z \simeq 3$ (Lowenthal et al. 1997; Franx et al. 1997;
Pettini et al. 1998; Pettini et al. 2000).
For the purpose of this paper, we are interested in establishing 
the maximum galactic mass which allows blowout 
to occur. MacLow \& Ferrara (1999) have shown that when
blowout does take place the escape efficiency of the 
metals produced by the supernova 
progenitors is close to unity. 

\subsection{Conditions for Blowout}

The blowout condition can be derived by comparing two characteristic
velocities: the blowout velocity, $v_b$, and the escape velocity of the
galaxy, $v_e$.  
To calculate these two velocities we start by defining a protogalaxy
as a two component system consisting of a dark matter halo and a gaseous disk.
We assume a modified isothermal halo density profile 
$\rho_h(r)=\rho_c/[1+(r/r_a)^2]$ 
extending out to a radius 
$r_{200}\equiv r_h = [3M_h/4\pi (200 \rho_{crit})]^{1/3}$, 
defined as the radius within which the mean 
dark matter density is 200 times the critical density 
$\rho_{crit}= 3 H_0^2(1+z)^3/8\pi G$
at the redshift $z$ 
when the halo is identified.
$M_h$ is the halo mass and $H_0= 100 h$~km~s$^{-1}$~Mpc$^{-1}$ 
is the present-day Hubble constant;
throughout this paper we adopt $\Omega=1$.

For such a halo the escape velocity 
can be written as 
\begin{equation}
\label{ve1}
v_e^2\simeq 4\pi p G \rho_c r_a^2  = {2 p G M_h \over r_h}, 
\end{equation}
with $p=1.65$ (MacLow \& Ferrara 1999). 
Note that $v_e = \sqrt{2p}~v_c$, where $v_c$ is the circular velocity of the halo.

Due to the dissipative nature of the gas, the baryons which initially should be 
distributed approximately as the dark matter, will lose pressure and collapse
in the gravitational field of the dark matter. The density ratio between these
two components is assumed here to be equal to its cosmological value, and therefore
the initial density of the gas in the protogalaxy is $\rho_g = \Omega_b \rho_h$.
If the halo is rotating, the gas will collapse in a centrifugally supported disk.
The radius of the disk can be estimated by imposing that the specific angular
momentum of the disk, $j_d$, is equal to that of the halo, 
$j_h$ (Mo \etal 1998, Weil \etal 1998) 
\begin{equation}
\label{jd}
j_h = \sqrt{2} \lambda v_c r_h = 2 v_c \ell_d = j_d
\end{equation}
where $\ell_d$ is the disk scale length and $\lambda$ the standard halo spin 
parameter. 
As shown by numerical simulations (Barnes \& Efstathiou 1987; 
Steinmetz \& Bartelmann 1995) $\lambda$ depends very weakly on 
$M_h$ and on the density fluctuation spectrum; its distribution is approximately
log-normal and peaks around $\lambda=0.04$. From eq. \ref{jd} we then obtain 
$\ell_d = (\lambda/\sqrt{2}) r_h$. For an exponential disk, implicitly assumed in
deriving eq. \ref{jd} above,  
the optical radius (\ie the radius encompassing 83\%
of the total integrated light) is $3.2\times \ell_d$. 
Since galaxies typically extend $\approx 2$ times
their optical radius in H~I (Salpeter \& Hoffman 1996), 
we adopt for the radius of the gaseous disk
\begin{equation}
\label{rd}
r_d = 4.5 \lambda r_h. 
\end{equation}
We now evaluate the scale height of the gas in the disk. This is
roughly given by
\begin{equation}
\label{H}
H \simeq  {c_s^2 r_d^2\over G M_h}\simeq 70 \lambda^2 \left({c_s\over v_e}\right)^2 r_h
\end{equation}
where we have used the expression for $r_d$ above; $c_s$ is the effective 
gas sound speed which also includes a possible turbulent contribution.
Note that 
\begin{equation}
\label{hsur}
{H\over r_d} = 15.3 \lambda \left({c_s\over v_e}\right)^2
\end{equation}
If all baryons bound to the dark matter halo were able to collapse, 
the mean gas density in the disk would be
\begin{equation}
\label{dd}
\rho_d \approx \left({r_h\over r_d}\right)^3 \left({r_d \over H}\right) \rho_g.
\end{equation}
The previous assumption, although uncertain, gives approximately the correct
density when scaled to Milky Way parameters; in addition, $\rho_d$ enters 
in the expression for $v_b$ below to the 1/3 power. Thus, unless the fraction
of collapsed baryons is unreasonably low, a value lower than unity does not
lead to qualitatively different conclusions.
The explicit expression for the blowout velocity, $v_b$,
has been obtained by Ferrara \& Tolstoy (2000):
\begin{equation}
\label{vb} 
v_b = 2.7 \left({L\over \rho_d H^2}\right)^{1/3},
\end{equation}
where $L$ is the mechanical luminosity of the parent SB.
Note that $v_b$ by definition (Ferrara \& Tolstoy 2000) is estimated at a height equal to 3H above
the galactic plane. Obviously, if the blowout velocity has already
decreased below the local sound speed at that point, the blowout is inhibited.             
By using eqs. \ref{rd}, \ref{dd} and \ref{H} it is easy to show that 
$H^2 \rho_d \propto \Omega_b c_s^2 $,
that is the product $H^2 \rho_d$ is independent of
mass and redshift; 
therefore $v_b$ depends on these quantities only through $L$.
Blowout occurs if $v_b > v_e$, or
\begin{equation}
\label{cond} 
2.7 \left({L\over H^2 \rho_d}\right)^{1/3} 
\ge \left({2 p G M_h \over r_h}\right)^{1/2}.
\end{equation}
This implies that the mechanical luminosity for a SB to blowout must be
larger than the critical value
\begin{equation}
\label{Lcrit} 
L_c = (0.27 p G)^{3/2} \rho_d H^2 \left({M_h\over r_h}\right)^{3/2}. 
\end{equation}
The {\it total} mechanical luminosity of the corresponding SB          
in a primordial galaxy that virialized 
at redshift $z$ can be written as (Ferrara 1998)
\begin{equation}
\label{L}
L_t(z)={\epsilon_0 \nu\dot M_\star}={\epsilon_0\nu \Omega_b f_b \over\tau_\star t_{ff}(z)}M_h,
\end{equation}
where $\epsilon_0=10^{51}$~ergs is the energy of a SN explosion and
$\dot M_\star$ is the star formation rate;
we assume a Salpeter IMF, for which one SN       
is produced for each 56~$M_\odot=\nu^{-1}$ of stars formed.
The baryon density parameter is $\Omega_{b}=0.05 \Omega_{b,5}$, of
which a fraction $f_b \sim 0.08 f_{b,8}$ (Ciardi \etal 2000a) is able to cool
and become available to form stars. 
The free-fall time is $t_{ff}= (4\pi G \rho_h)^{-1/2}$;
$\tau_\star^{-1}=3\%$ is the star formation efficiency, which is fixed by
matching to the cosmic star formation history as described in \S4.2 below. 

For Population III objects, which
as we have argued are likely to have been of low mass,
we make the simplifying assumption that star formation is confined to a 
relatively small region, 
and that all the exploding SNae associated with the initial star 
formation episode result in the formation of a single superbubble.
In this case a comparison between eq. \ref{Lcrit} and eq. \ref{L} 
shows that $L_t(z)/L_c$ does not depend on mass and redshift,
and the condition $L_t(z)>L_c$ is satisfied when
\begin{equation}
{\epsilon_0 \nu f_b\over \tau}> 0.1 c_s^2,
\end{equation}
which for typical parameters is equivalent to
$4.4\times 10^{12}>0.1 c_s^2$, and is therefore valid for
galaxies with $c_s\leq 70$ km/s. 
Thus low mass galaxies
with coherent star formation (\ie with all SNae driving a single SB)
are very likely blowing out.

For more massive and larger galaxies 
we have to consider the more realistic  
situation where the SNae are more widely distributed
within the disk, and occur in 
OB associations with different values of $L_{OB}$
(or, equivalently, with a different number $N$ of SNae). 
In nearby galaxies it is found that the luminosity function of 
OB associations is well approximated by a power-law
\begin{equation}
\label{dist}
\phi(N) = {d{\cal N}_{OB}\over dN} = A N^{-\beta}
\end{equation}
with $\beta \approx 2$ (McKee \& Williams 1997; Oey \& Clarke 1998). 
Here ${\cal N}_{OB}$ is the number of associations containing
$N$ OB stars; 
normalization of $\phi(N)$ to unity requires
$A=1$. 
Thus the probability for a cluster of OB stars
to host $N$ SNae is $\propto N^{-2}$,
where $N=L_{OB}t_{OB}/\epsilon_0$, and $t_{OB}=40$~Myr 
is the time at which the lowest mass
($\approx 8 M_\odot$) SN progenitors expire. 
The total mechanical luminosity, 
which must be equal to $L_t(z)$ in eq.~\ref{L},
is then found to be 
\be
L_t(z)=\int\limits_{N_m}^{N_M} L_{OB}(N)\phi dN,
\ee
where $N_m=1$ ($N_M$) is the minimum (maximum) possible number of SNae in a cluster.
This gives
\be
L_t(z)={\rm const.}~{\epsilon_0\over t_{OB}}{\rm ln} {N_M\over N_m}.
\ee
The contribution to the total luminosity 
from clusters powerful enough
to lead to blowout is
\be
L_{B}(z,>L_c)={\rm const.}~{\epsilon_0\over t_{OB}} {\rm ln}{N_M\over N_c},
\ee
where $N_c$ is the number of SNae in a cluster with mechanical
luminosity equal to $L_c$, i.e.
\be
N_c={L_c t_{OB}\over \epsilon_0}.
\ee
Thus, the fraction of the mechanical energy 
which can be blown out is
\be
\delta_{B}={{\rm ln} (N_M/N_c)\over {\rm ln} (N_M/N_m)}<1.
\ee
Clearly, $N_M$ (and therefore $\delta_{B}$) is an intrinsically 
stochastic number. To determine its dependence on the total number of 
supernovae $N_t=L_t(z) t_{OB}/\epsilon_0$ produced by a galaxy 
during the lifetime of an OB association, 
we have used a Monte Carlo procedure
applied to the distribution function in eq. \ref{dist}. 
The results
for $N_M$ and $\delta_{B}$ as a function of $N_t$ are shown in 
Fig. \ref{fig1}; we recall that the star formation
rate $\dot M_\star$ in the galaxy is related (through eq. \ref{L}) 
to $N_t$ by $\dot M_\star \approx 5\times 10^{-6} N_t
~M_\odot$~yr$^{-1}$. 
As can be seen from Fig. \ref{fig1},
for low values of $N_t$ the quantity
$N_M$ is larger than $N_c$, implying that
in every galaxy at least some SBs are able to blow out.
However, near $N_t=10^4$ $N_M$
flattens and eventually becomes equal to $N_c$ at $N_t \simeq 45\,000$.
Above this limit (corresponding to a galaxy with $\dot M_\star \approx 
0.35~M_\odot$~yr$^{-1}$ or $M_h \approx 10^{12}(1+z)^{-3/2} M_\odot$) 
blowout is inhibited.  
The fraction $\delta_{B}$ can be seen from Fig. \ref{fig1}
to be a decreasing function of $N_t$; 
an approximate analytical form is
\ba
\label{fit}
\delta_{B}(N_t) = 1   {\rm ~~for~} N_t < 100 \\ \nonumber
\delta_{B}(N_t) = a+b{\rm ~ln} (N_t^{-1})   {\rm ~~for~} N_t > 100,
\ea
with $a=1.76$, $b=0.165$. 
Clearly, in small galaxies even the smallest associations
are capable of producing blowout, 
so that the issue of coherence discussed above
is irrelevant.

Thus, if we assume the most of the baryonic matter has collapsed into the disk,
the main metal polluters of the IGM 
at the present epoch are galaxies with visible mass lower than 
$M_d^{up} =\Omega_b M_h \simlt 5 \times 10^{10} M_\odot$. 
This critical mass decreases with increasing
redshift as $(1+z)^{-3/2}$, indicating
that the initial metal enrichment of the universe 
must have been produced by even smaller
galaxies. 
For example, at $z = 20$, the expected value of $M_d^{up}$ is only $5\times 
10^8 M_\odot$. 

\subsection{Confinement of Blowout-driven Outflows}

Blowout will drive an outflow which will 
eventually be confined by the
IGM pressure. 
What is the characteristic length, $R_e$, 
at which such pressure equilibrium is
achieved? 
By requiring that the outflow ram pressure, $\rho_w v_w^2$, is equal to the
IGM pressure, $p_{i}$ we obtain
\begin{equation}
\label{Re} 
R_e = \left(\dot M_w v_w \over 4\pi p_i\right)^{1/2};
\end{equation}
where we have used mass conservation 
and assumed that the flow is approximately spherical.
Once recombination is complete at redshift $z\approx 200$, 
and before the IGM is reheated by
the energy input from the first galaxies, 
the IGM pressure will evolve almost adiabatically,
$p_{i}=p_\star[(1+z)/200)]^5$, 
where $p_\star/k_B\approx 1500 h^{2}$~cm$^{-3}$~K is the value of 
$p_i$ at $1+z=200$. 
However, almost unavoidably, even the earliest
SN-driven bubbles will expand in an IGM which
has been pre-ionized by the same massive stars which later exploded as SNae. 
Thus, it seems more appropriate to calculate the IGM pressure 
in the surroundings of the 
galaxy as $p_{i}= (R/\mu)\rho_i(z) T$, 
with $T\approx 2\times 10^4$~K as a result of photoionization heating.
This is a reasonable hypothesis since it can be shown 
Ciardi \etal 2000a) that the ionization spheres are
much larger than the metal-enriched bubbles considered here. 
The mass loss rate $\dot M_w$ is typically a fraction $\xi \approx 10$\% of the 
star formation 
rate $\dot M_* =L_e/\epsilon_0 \nu$ 
(Ferrara \& Tolstoy 2000) where $L_e=\delta_{B}L_t$ is
the effective mechanical luminosity, that is the fraction available for blowout; 
we can also assume that $v_w \simeq v_e$. With these
hypotheses, and using the relations derived above for $v_e$, one obtains 
\begin{equation}
\label{Re1} 
R_e(M_h, z) = 0.9 \left({\Omega_{b,5}\over \lambda_4}
{v_e^4 \delta_{B}\over p_{i}}\right)^{1/2} {\rm ~cm},
\end{equation}
where $\lambda_4=\lambda/0.04$.
The behavior of $R_e$ with $M_h$ is shown in Fig. \ref{fig2} for different redshifts.
At lower redshifts, the bubbles are larger because the pressure of the 
confining IGM is lower; also evident from the figure 
is the increase of the critical mass for blowout with decreasing $z$.

The major conclusion that can be drawn from Fig. \ref{fig2}
is that supernova-driven outflows are quite inadequate for
dispersing heavy elements far from their production sites and 
cannot account, by themselves, for the ubiquitous 
presence of metals in the IGM at $z \simeq 3$.  
This can be readily realized when we consider that 
in standard Cold Dark Matter models, taken here as representative of a larger 
class of hierarchical models of structure formation, 
the typical (physical) separation
between objects with mass $M_h\approx 10^{10} M_\odot$ 
varies between $2-0.2$~Mpc 
in the interval $0 < z < 8$ (Ciardi \etal 2000a). 
This scale is much larger than the values of 
$R_e$ in Fig. \ref{fig2}, so that we would expect regions outside the metal 
enriched bubbles to retain their primordial composition 
(this argument is reconsidered more rigorously in \S4 below).
Clearly some other mechanism must be operating to remove the metals  
from the surroundings of galaxies and mix them with the more generally 
distributed IGM.

\section{Predictions for CDM Models}

To make further progress, we need to fix a specific cosmological
model. As an example, we consider the so-called Standard Cold Dark
Matter (SCDM), with $\Omega_M=1, \Omega_\Lambda=0$, and $h=0.5$;
the power spectrum $\vert
\delta_{k} \vert^{2}$ is taken from Efstathiou \etal (1992),
normalized to the present-day abundance of rich clusters
($\sigma_8 =0.6$; Eke, Cole, \& Frenk 1996).
To calculate the number density of dark matter halos 
as a function of redshift we use the 
Press \& Schechter (1974, hereafter PS) formalism; this
technique is widely used in semi-analytical models of galaxy formation
and gravitational lensing
(White \& Frenk 1991, Kauffman 1995, Ciardi \&
Ferrara 1997, Baugh \etal 1998, Guiderdoni \etal 1998, Marri \& Ferrara 1998) and it
has been shown to be in surprisingly good
agreement with the results from N-body numerical simulations.
Given a power spectrum $\vert \delta_{k} \vert ^{2}$, 
one can write the Gaussian variance of the fluctuations on the mass scale $M$:
\begin{equation}
\sigma_{M}^{2}=\int
\frac{d^{3}k}{(2\pi)^{3}}W^{2}(k,R)|\delta_{k}|^{2},
\label{sigmam}
\end{equation}
where $M=(4/3)\pi \rho R^3$, $\rho$ is the matter density, and
\begin{equation}
W={3\over (kR)^3} \left[ \sin (kR)-(kR)\cos (kR) \right];
\label{wf}
\end{equation}
is a top-hat filter function.  From the results
of the nonlinear theory of gravitational collapse,
stating that a spherical perturbation with overdensity $\delta_c = \delta \rho/
\rho > 1.69$ with respect to the background matter collapses to form a
bound object, through the PS formalism we can derive the normalized fraction
of collapsed objects per unit mass at a given redshift:
\begin{equation}
f(M ,z)=\sqrt{\frac{2}{\pi}}
\frac{\delta_{c}(1+z)}{\sigma_{M}^{2}}
e^{-\delta_{c}^{2}(1+z)^2/2\sigma_{M}^{2}}
\left( -\frac{d\sigma_{M}}{dM} \right).
\label{fmz}
\end{equation}
Then the comoving number density of dark matter halos per unit mass is       
\begin{equation}
n_h(M_h,z)={\Omega_M \rho_{crit}\over M}f(M_h ,z).
\label{nh}
\end{equation}
We can now ask what is the fraction of the IGM polluted by metals at different
redshifts. To this end we calculate an IGM porosity parameter, $Q(z)$, defined by
\be
\label{Qz}
dQ(z) = M_h\vert {dn_h\over dz}\vert R^3(M_h,z) dz,
\ee
in two different cases. First, we consider the case in which metals are only
injected in the IGM by superbubbles 
and therefore $R\equiv R_e$ in the previous equation.
The resulting porosity evolution 
(complete overlap of the bubbles occurs for $Q=0.16$, Smith 1976) is shown in
Fig. \ref{fig3}. 
Again we see that blowout by itself would
lead only to a negligible dispersal of metals ($Q < 10^{-4}$),
with most of the IGM maintaining its primordial composition to the 
present day. 

If we are to explain the relatively ubiquitous presence of metals
in the `true' IGM, as deduced from at least some observations, 
we are then forced to assume that some additional physical
mechanism, capable of efficiently transporting metals away from
their production sites, must be at work. The nature of such a
process can only be matter of speculation at present, because
neither the numerical simulations nor the observations have yet
reached the required levels of sophistication or sensitivity to
address this question properly. Among the options which are
worth exploring are galaxy collisions, diffusive processes and
the peculiar motions of galaxies. We consider the relative
importance of these different possibilities in future work.

Here we take a strictly phenomenological approach which
nevertheless has considerable predicting power. We introduce a
diffusive radius, $R_d$, defined as the mean interdistance
between the galaxies responsible for providing the predominant
contribution to the metal enrichment of the IGM (see Fig.
\ref{fig4}), \ie $R_d=0.2$ Mpc (comoving) for $M_h > 2\times 10^8
M_\odot$. As the largest Doppler parameters measured in
Ly$\alpha$ clouds are of order $\approx 50$~km~s$^{-1}$, the time
required for a pollution front to travel such a distance is
shorter than the Hubble time only at redshifts $z<1$. Thus, our
assumption is equivalent to fixing at $z \simlt 1$ the epoch at
which metals become homogeneously distributed. As we will see
shortly, this simple hypothesis leads to a number of consequences
which are in accord with available data. If future studies of 
possible mechanisms for the diffusion of metals are able to
determine directly the value of $R_d$ and its evolution with
time, it will be relatively easy to include their results in
the general framework of this paper and explore any differences
with the conclusions presented here.

As can be seen by Fig. \ref{fig3}, when the condition
$R=max(R_e,R_d)$ is introduced, mixing is improved
dramatically---as expected, and metal-enriched bubbles indeed
overlap at $z\approx 1$. At later epochs essentially all of the
IGM is polluted with heavy elements produced in galaxies and
subsequently redistributed by the combined effects of blowout and
diffusion.

\subsection{Metallicity of Polluted Regions}    

We can now calculate the average metallicity of the gas inside the diffusive
spheres. In order to do so we need to know $\mu_Z$, the mass of metals 
produced by the typical supernova. Nucleosynthesis
calculations in Type II SNae by Tsujimoto \etal (1995) predict
$\mu_Z \approx 2 - 3 M_\odot$ for a Salpeter IMF, depending  
on the upper mass limit above which a black hole is formed.
Here we adopt $\mu_Z=2.58 M_\odot$, also consistent with a   
matching to the cosmic star formation history (\S4.2).
By design, mixing is
efficient and the metal distribution is therefore homogeneous 
within the radius $R_d$. 
The total number of SNae per galaxy is 
\be
{\cal N}_{SN}={\nu\Omega_b f_b\over \tau_\star}M_h.
\ee
Then the mass of metals ejected by a galaxy inside a halo of mass $M_h$ is 
\be
M_e={\mu_Z\nu\Omega_b f_b \over \tau_\star}M_h\delta_{B},
\ee
with $\delta_{B}$ given by eq. \ref{fit}. 
The metal density in a diffusive sphere surrounding a given halo 
is $\rho_Z=3M_e/4\pi R_d^3$, 
and its average metallicity is $\rho_Z/\rho_i(z)$, where $\rho_i(z)$
is the mean density of IGM at redshift $z$.
This assumption restricts our analysis to the case of the `true' IGM,
characterized by overdensities close to unity. Extrapolation of the results
to hydrogen column densities $\log N_{HI} \simgt 14$ is only very qualitative
and deserves more study.
The dependence of metallicity in the diffusive spheres on halo mass and
redshift is plotted in Fig. \ref{fig4}. 
At any given redshift, $Z$ increases
as a function of mass (because of the increasing metal production) up to an
abrupt cutoff when blowout becomes inhibited, as described by
the behavior of $\delta_{B}$ in Fig. \ref{fig1}. 
As the spheres grow with time, the metals are distributed 
over larger volumes and this is reflected by the 
shift to lower metallicities with decreasing $z$ in Fig.~{\ref{fig4}.
Note that for the same reason pockets of very high metal content
($Z\approx 1 Z_\odot$) are expected at large $z$, although
the size of these regions is very small.

\subsection{IGM Metallicity Evolution}    

Once the cosmological model has been
fixed, we need to determine the values of the 
the star formation efficiency, $\tau_\star$,  
in order to calculate the metallicity of the IGM
as a function of time. We choose this  value
by comparison with observations, as follows. 
We calculate the evolution of $\Omega_\star$, the density parameter of 
stars formed in the universe (see Fig. \ref{fig5}),
with the requirement that at $z = 0$ it matches 
the estimate by Fukugita, Hogan \& Peebles (1998).
These authors conclude that $\Omega_\star(0)$ in spheroids, disks and irregulars is
$\approx 0.0049 h^2_{50}$. 
A similar value is found by integrating  
current estimates of the  
star formation rate density in the universe as a function of redshift,
as deduced from deep imaging surveys (Pettini 1999).
With this normalization (giving $\tau_\star^{-1} \approx 3\%$), we can 
then derive the corresponding evolution
of the metals produced by stars and returned to the ISM
$\Omega_Z$ (Fig. \ref{fig5}), which is directly proportional $\Omega_\star$. 

However, not all the metals can escape
from the galaxy where they have been produced. 
The cosmic ejection fraction, $f_{ej}(z)$,
(\ie the ejection fraction averaged over the entire population of halos) 
is very close to unity
at high redshift where predominantly small galaxies are present, 
but it steadily 
decreases to about 50\% at $z=0$, 
as the number of more massive galaxies able to retain
their metals increases. 
For this reason, the curve describing the density parameter 
of {\it ejected} metals $\Omega_Z^{ej} =  f_{ej}(z)\Omega_Z(z)$ 
in Figure 5 progressively deviates 
from that for $\Omega_Z(z)$ with decreasing redshift.  
Stated differently, today about 50\%  of the
metals produced should reside in the IGM. 
If the metals were homogeneously mixed with 
the baryons in the universe at any redshift 
the average IGM metallicity (top  
curve in Fig. \ref{fig5}) 
would be $\langle Z \rangle=\Omega_Z^{ej}/\Omega_b \simeq 1/25 
Z_\odot$ at $z = 3$ and
$\langle Z \rangle \simeq 0.1 Z_\odot$ at $z = 0$. 
Note that the average metallicity of today's galaxies
would be higher by a factor $\Omega_b/(\Omega_\star + \Omega_g)$
(the baryon density 
divided by the sum of star and gas density in galaxies);
we therefore expect that for luminous matter today 
$\langle Z \rangle \approx  Z_\odot$, 
in accord with observational 
estimates (Edmunds \& Phillips 1997).

Our main results are displayed in Figure 6 and 7, 
which show in the spread in metallicity as a function of redshift,
compared with the average $\langle Z \rangle$ of the IGM
from Figure 5.
The two figures correspond to different metallicities of the outflowing 
hot gas in the superbubbles, assumed to be $Z_{\odot}$ in Fig. 6 and
$8 Z_{\odot}$ in Fig 7.
In the figures the metallicity distribution at each redshift considered
is shown in the vertical direction with the density of symbols
approximately proportional to the amplitude of the distribution.
The distributions are relatively flat, but at redshifts 
$z \simlt 5$ they also show a double-horned profile 
with points accumulating at the highest and lowest metallicities.
We illustrate this effect in Figure 8, where we have 
reproduced the metallicity histograms at $z = 0$ and $z = 1$
from Figure 6. The more massive galaxies are responsible
for the high metallicity peaks in these distributions, caused by the 
cutoff in the ejection fraction described by the parameter 
$\delta_B$ (Figure 1). The low metallicity maximum is due to the 
increasing number of low mass halos.

There are several interesting features of Figures 6 and 7
which we now discuss briefly.
First, at $z > 6$ the average IGM metallicity 
is not within the predicted ranges of metallicities, 
being lower than the lowest values in the distributions.
What we are seeing is a highly inhomogeneous distribution of metals
which are still clumped in small regions around
the galaxies which produced them.
As mixing proceeds, the average metallicity 
increasingly becomes a better description of the true mean metallicity,
and approaches the mean of the distribution.  
At $ z \approx 1$ the postulated diffusion leads to 
all the volume in the universe being polluted with heavy elements
to some extent
(Figure 3). The merging of the metal-enriched spheres produced by 
different galaxies  has the effect of erasing regions of 
low metallicity: the minimum metallicity corresponds to that 
in spheres marginally overlapping at a given redshift. 
This effect is reflected in the
growth of the lower boundary of the distributions
from  $(1+z) \simlt 5$ in Figures 6 and 7,
and results in the present-day IGM metallicity being confined within the 
narrow range $0.1 \pm 0.03 Z_\odot$(1 $\sigma$). 

The decreasing spread in metallicity with time is a direct result
of the growth of $\langle Z \rangle$. At high $z$, where 
the overall level of enrichment is low, 
even relatively metal-poor spheres have a chance to stand out.
However, as the average IGM metallicity increases, 
only a few massive objects are
able to produce diffusive spheres with $Z$ sufficiently high 
to be recognized as metallicity
peaks (and still be able to eject their metals).
The situation is similar
to a ``flooding effect'',  where as the water level rises  
fewer and fewer mountain peaks can be seen.

At redshifts $z > 1$, when metal enrichment is highly
inhomogeneous, a considerable fraction of the IGM is still of
essentially primordial composition. Thus, at $z \approx 3$ for
example, we expect that only some low column density Ly$\alpha$
forest clouds will show associated metal lines, while the
majority will not; the ratio between Ly$\alpha$ clouds with and
without metals depends on the covering factor of the diffusive
spheres and grows with time. This has implications for the
interpretation of the results of searches for metals in the forest.
Given the low optical depths when log~$N$(H~I)~$< 14$, such
searches are normally conducted in a statistical way, by
considering together the data from many absorption lines. In such
cases one obtains some gross average over all the 
absorbers which, given the dilution with the unpolluted IGM, will
be systematically lower than the values of $\langle Z \rangle$
plotted in Figures 6 and 7. For this reason we also show in these
figures the covering factor-weighted metallicity 
$\langle Z \,P(z) \rangle$, 
where $P(z)$ describes the evolution of the covering factor 
down to $z = 1$\,.

In Figures 6 and 7 we have used different symbols to represent the contributions
to the IGM metallicity from 
halos with virial temperatures above (hexagons) and below (triangles) 
$10^4$~K, to give a qualitative idea of the role of large and small objects
in the enrichment process. It is seen that at high $z$ small objects are 
controlling the process, whereas at lower redshift enrichment is largely regulated by 
more massive galaxies (up to the ejection limit set by $\delta_{B}$, see
Figure \ref{fig3}). If some inhibiting effect, such as photoheating 
by the UV background, affects preferentially low mass galaxies, 
the lower metallicity bounds
at high redshift will move up accordingly, shifting closer to the 
line separating hexagons from triangles.     
The flat upper boundary of the distribution at high $z$ reflects the
fact that the volumes involved are so small that the IGM baryon loading 
results in a negligible dilution of the metallicity of the ejecta.
Since there is no firm measurement of this quantity at present 
(Heckman, private communication), we have considered
two possibilities, $Z_\odot$ (our standard case) and 8 $Z_\odot$
in Figures 6 and 7 respectively.
The IGM metallicity distribution is affected by this
choice at high redshift, but the difference becomes much smaller at low $z$,
where the IGM baryon loading regulates 
the dilution of the larger diffusive spheres.

Finally, we also show in the two figures 
the average metallicities for the case in which 
metal-enriched spheres with size below $R_e(min) = 1$~kpc have been 
excluded. In our models these are 
the absolute lower limits to the average IGM metallicity at a given redshift.
$R_e(min)$ has been calculated by imposing the condition 
that the collisional timescale 
between galaxies with a given impact parameter is shorter than the Hubble time
at any redshift up to $z=10$. Obviously, a sphere with size greater than
$R_e(min)$ can 
be strongly disturbed by tidal interactions following an encounter at greater
impact parameter, and for this reason the curves obtained 
in this way are strictly lower limits.
Nevertheless, as can be seen from Figure 6 and 7,
imposing this condition reduces the metallicity by only
about 40\%.

In addition to the SCDM model, we have also considered a CDM model with a
cosmological constant (CDM+$\Lambda$) with $\Omega_M=0.4, \Omega_\Lambda=0.6, 
h=0.5$). By normalizing the star formation rate following the same 
procedure as for the SCDM model we obtain a higher value of $\tau_\star^{-1}
=10.5\%$. Because of the normalization, the resulting
metal distribution is qualitatively very similar to the one derived for
the SCDM case. The only notable
differences are a larger metallicity spread between $0.5 < z < 2$ and a
somewhat higher mean value at $z=0$,  
$\langle Z \rangle = 0.15 \pm 0.03 Z_\odot$.

\section{Discussion}    

The basic conclusion of this paper is that metal ejection driven
by SN events fails, by more than one order of magnitude, to
distribute the products of stellar nucleosynthesis over volumes
large enough to pollute the whole IGM to the typical metallicity
of \lya\ clouds, [C/H]~$ \simeq -2.5$. We are therefore forced to
conclude that some additional physical process must be at play,
the nature of which remains to be determined.
In our scheme transport of metals occurs in two sequential steps.
First, SNae provide the initial kick necessary to eject heavy
elements outside the potential well of a galaxy, but the ejecta 
are then confined by the IGM pressure to a relatively small
bubble, of radius $R_e$. We then postulate that a second process
is responsible for the diffusion of metals on a typical scale
$R_d$, comparable to the mean distance between the galaxies which
are the most efficient pollutants of the IGM. 

Our scenario is qualitatively different from that proposed by
Gnedin (1998), who attributed the mixing to more violent and
rarer galaxy mergers. In the model proposed here SNae are of
fundamental importance as they initiate the mixing process.
However, galaxy collisions might well play a role in the
subsequent phase during which metals are spread over larger
scales, of order $R_d$. Other diffusive processes, such as
thermal conduction and turbulent mixing layers at the interfaces
between cosmological flows, as well as the peculiar motions of
galaxies, may also be important in determining the spatial
structure of the distribution of metals. A detailed study of such
processes is beyond the scope of this paper and should, in any
case, be based on cosmological simulations which are in progress.
We will report of this extension of the work in a forthcoming
paper.

In the present study we have stressed that the IGM enrichment
proceeds in a very inhomogeneous manner, with pockets of metal
rich gas gradually increasing both in number and in size until
they overlap at $z\approx 1$. The average metallicity of the IGM
increases with time, a trend confirmed by the results in Barlow
\& Tytler (1998), who found an order of magnitude increase in the
metallicity of the Ly$\alpha$ forest between $z=2.5$ and $z=0.5$.
Although our results are strictly applicable only to the `true'
IGM with overdensities close to unity, we nevertheless regard
this as an encouraging performance of our model. 

The metallicity spread is predicted to decrease with the progress
of time. At $z \simlt 1$, when the entire volume of the universe
has been exposed to metal pollution, the spread in metallicity is
less than one order of magnitude, and at the present epoch 
$\langle Z \rangle = 0.1 \pm 0.03 Z_\odot$. Thus we predict that
at $z < 1$ essentially all absorbers should have associated C~IV
absorption, irrespectively of their column density. The effect
should be very pronounced. Not only are the voids polluted by the
overlap of metal-enriched diffusive spheres, as discussed above, 
but the decreasing intensity and hardness of the ionizing
background lead to a further increase in the fraction of C which
is triply ionized, so that the ratio $N$(C~IV)/$N$(H~I) increases
for a fixed [C/H] (Rauch et al. 1997). It should be possible to
test these predictions with forthcoming observations.
UV-efficient echelle spectrographs, now available on the VLT and
soon on the Gemini South telescope, will allow sensitive searches
for C~IV doublets at significantly lower redshifts than probed so
far, while STIS on {\it HST} and FUSE will map the \lya\ forest
with the required spectral resolution at wavelengths below
3000~\AA, which are inaccessible from the ground.

While our study has concentrated on the IGM, it also has
important consequences for the metallicity of the intracluster medium. 
which is found to have a rather uniform  value $\approx 1/3 Z_\odot$ 
with little, if any, evolution up
to $z \simeq 0.3$ (Fukazawa \etal 1998, Renzini 1999). 
This has been interpreted as
evidence supporting the view that 
the enrichment process in clusters was already completed 
by that epoch. 
In general we expect two sources to contribute to the build up of 
the intracluster medium, infalling IGM and gas stripped from cluster 
galaxies. In our models these two components have 
significantly different composition.
If the clusters formed at $z \simeq 1$,
we expect the IGM to have
$\approx 0.1 Z_\odot$ (roughly constant from 
$z \simeq 1$ to the present),
while the galaxies' ISM has 
approximately solar composition.
The simplest estimate of the resulting metallicity is
the geometric mean of these two values
(as appropriate for hydrodynamical mixing problems, see
Begelman \& Fabian 1990), that is  
$\langle Z \rangle_{icm}
= (0.1 \times 1)^{1/2} Z_\odot= 0.32 Z_\odot$, as observed. 
While this may well be a fortunate coincidence, it is also true that this 
conclusion lends qualitative support to the model and the assumptions made.

We make a final point concerning the role of intergalactic dust. 
The gas phase abundances we have derived 
do not take into account the possibility that some fraction of heavy
elements may be locked up into dust grains. 
If this were the case, clearly the distributions of metallicities in 
Figure 6 and 7 would shift towards lower values of $Z$.
In addition, at redshifts were the enrichment is still inhomogeneous,
dust associated with regions of high metal concentration may 
reprocess UV/optical photons into IR radiation and give rise
small scale anisotropies in the Cosmic Microwave Background which may be 
detectable (Ferrara \etal 1999). Having said this, however, we consider 
it highly speculative whether dust can survive at all in the hostile 
environments associated with outflowing superbubbles, where the 
temperatures are high and the gas has been shocked by SN explosions.

\section{Summary}    

In this paper we have investigated the evolution of the
metallicity  of the intergalactic medium with particular emphasis
on its spatial distribution. We have derived the conditions under
which supernova-driven ejection of metals from galaxies can
occur.  A strong conclusion of our calculations is that if SNae
were the only source of kinetic energy for the metals, a highly
inhomogeneous distribution would result at any redshift. Under
these conditions we would expect most of the volume of the
universe to remain at near-primordial composition, with a
metallicity $Z \approx 10^{-4} Z_{\odot}$, in contrast with the
observational results discussed in the Introduction. Thus, an
additional (but yet unknown) physical mechanism must be invoked
to mix the metals on scales comparable to the mean distance
between the galaxies which are the most efficient pollutants. From
this simple hypothesis we have derived a number of testable
predictions for the evolution of the IGM metallicity.

Quantitatively, we find that:

1. Metal ejection, or blowout, is inhibited in galaxies with 
total mass above $M_h \approx 10^{12}(1+z)^{-3/2} M_\odot$ due to
the combined effects of their larger gravitational field {\it and} 
less coherent SN energy deposition. 

2. The fraction of metals ejected over the star formation history of the 
universe is about 50\% at $z=0$. We expect that at the present epoch 
approximately half of the metals are to be found in the IGM and
the average metallicity of luminous matter to be approximately solar.

3. If the ejected metals were homogeneously mixed with 
the baryons in the universe, the average IGM metallicity 
would be $\langle Z \rangle=\Omega_Z^{ej}/\Omega_b \simeq 1/25 
Z_\odot$ at $z=3$. However, 
due to the spatial inhomogeneity,  
$\langle Z \rangle$ is actually lower than the mean of the distribution
in the diffusive metal-enriched spheres.

4. Metals become homogeneously distributed in the IGM at 
$z\simlt  1$, when the metal-enriched zones overlap,
and the spread of the distribution is reduced.
We calculate that at $z = 0$ the IGM metallicity is in the
range $Z \approx 0.1 \pm 0.03 Z_\odot$.

5. The uniform metal abundance of intracluster gas 
from $z \simeq 0.3$ to the present 
is naturally explained by a mixture of  
IGM infalling into the cluster and gas stripped from cluster members,
with element abundances for both components as predicted by our models.
\bigskip

We should like to thank E. Corbelli, A. Meiksin and B. Nath for 
useful discussions.  YS acknowledges support from the OAArcetri. 


\newpage

\newpage
\begin{figure}
\centerline{\psfig{figure=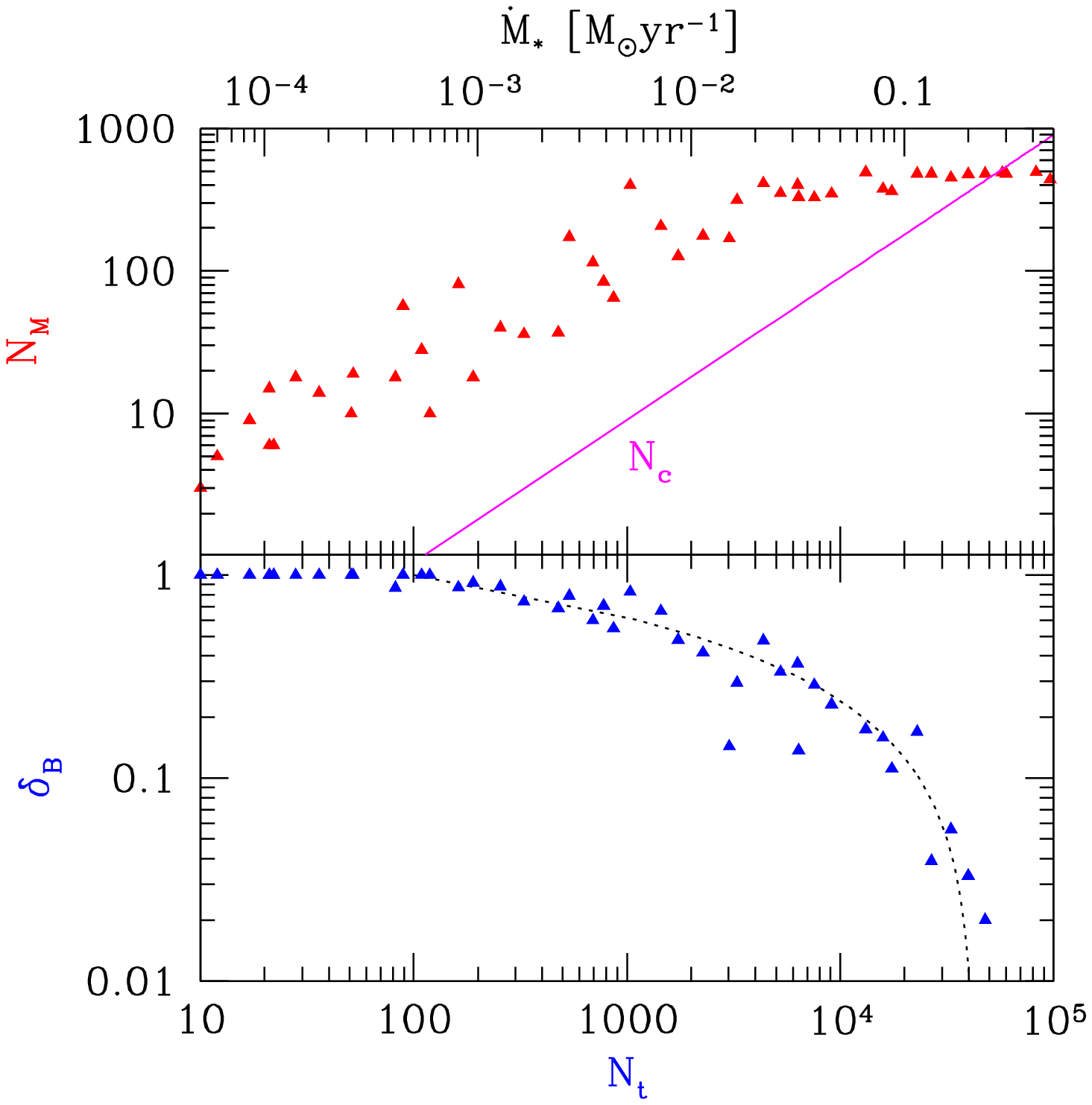,height=20cm}}
\caption{\label{fig1}\footnotesize{{\it Upper panel}: dependence of the maximum number of 
SNae in a
cluster as a function of the total number of SNae, $N_t$, produced by a given galaxy
in an OB association lifetime as obtained from  Monte Carlo simulations; 
also shown is the critical number of SNae, $N_c$, for blowout.
{\it Lower panel}: Energy fraction available for blowout, $\delta_{B}$ as 
a function of the total number of SNae produced by a given galaxy
in an OB association lifetime as obtained from  Monte Carlo simulations;  
also shown is an analytical fit to the numerical data.  
}}
\end{figure}

\newpage
\begin{figure}
\centerline{\psfig{figure=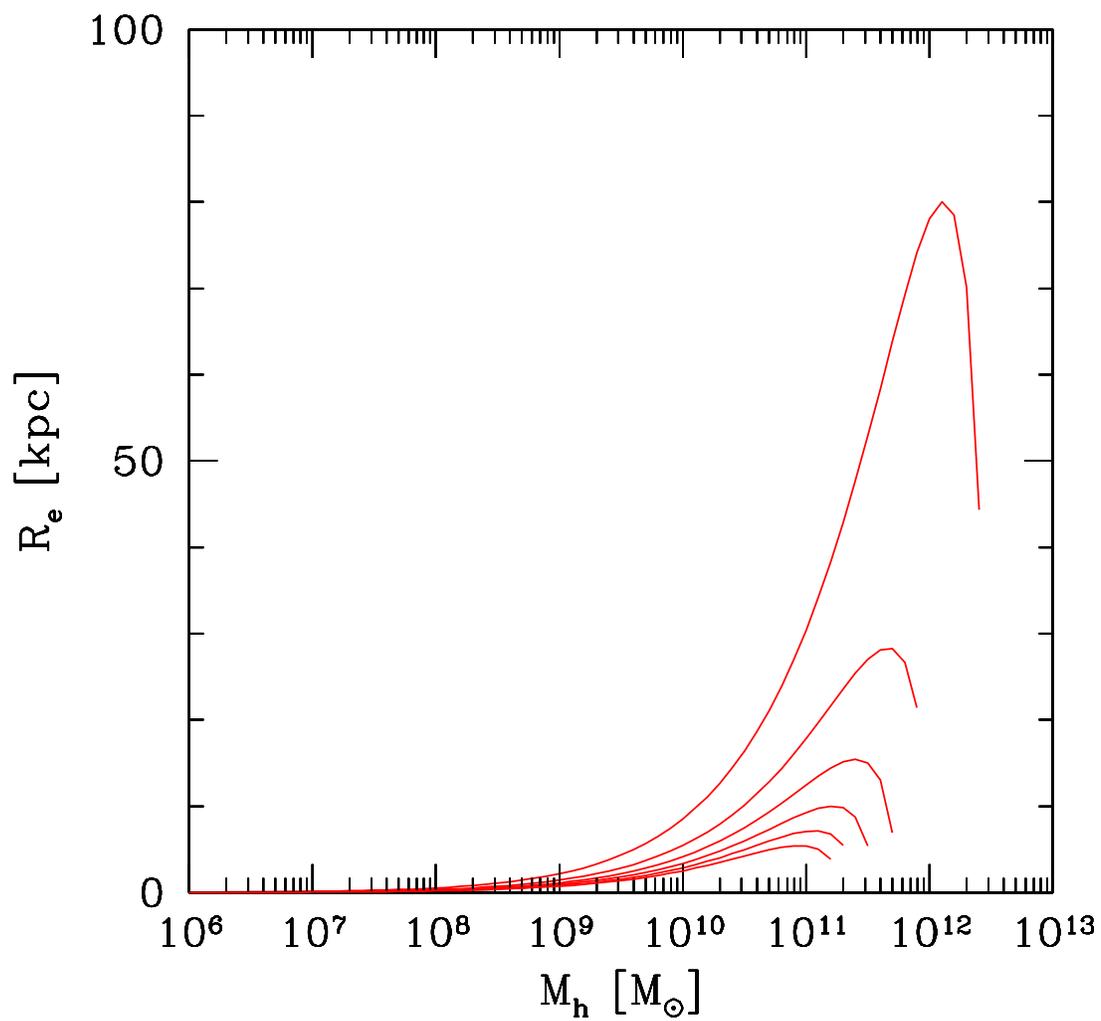,height=20cm}}
\caption{\label{fig2}\footnotesize{Radius of the metal bubble (in physical units) 
as a function of the halo mass for different virialization redshifts $z=0,1,2,3,4,5$ 
from the uppermost to the lowermost curve, respectively. The endpoint of each curve denotes
the critical mass for blowout.
}}
\end{figure}

\begin{figure}
\centerline{\psfig{figure=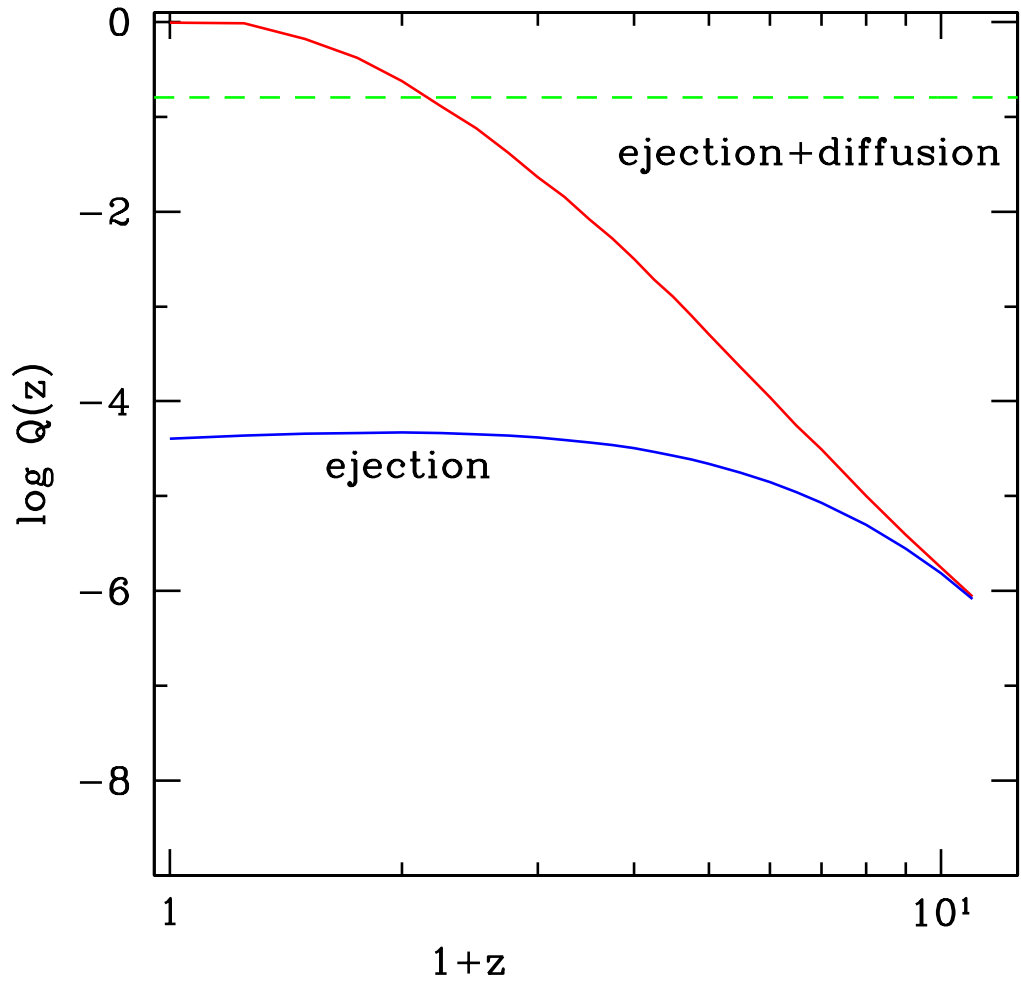,height=20cm}}
\caption{\label{fig3}\footnotesize{ Porosity factor of the metal bubbles considering
transport by ejection driven by superbubbles (lower curve) and transport due ejection
and diffusion processes acting in combination (upper curve). Overlapping of
the various bubbles is seen to occur at $z\approx 1$, where $Q(z)$ is equal to the
critical value 0.16.
}}
\end{figure}
\begin{figure}
\centerline{\psfig{figure=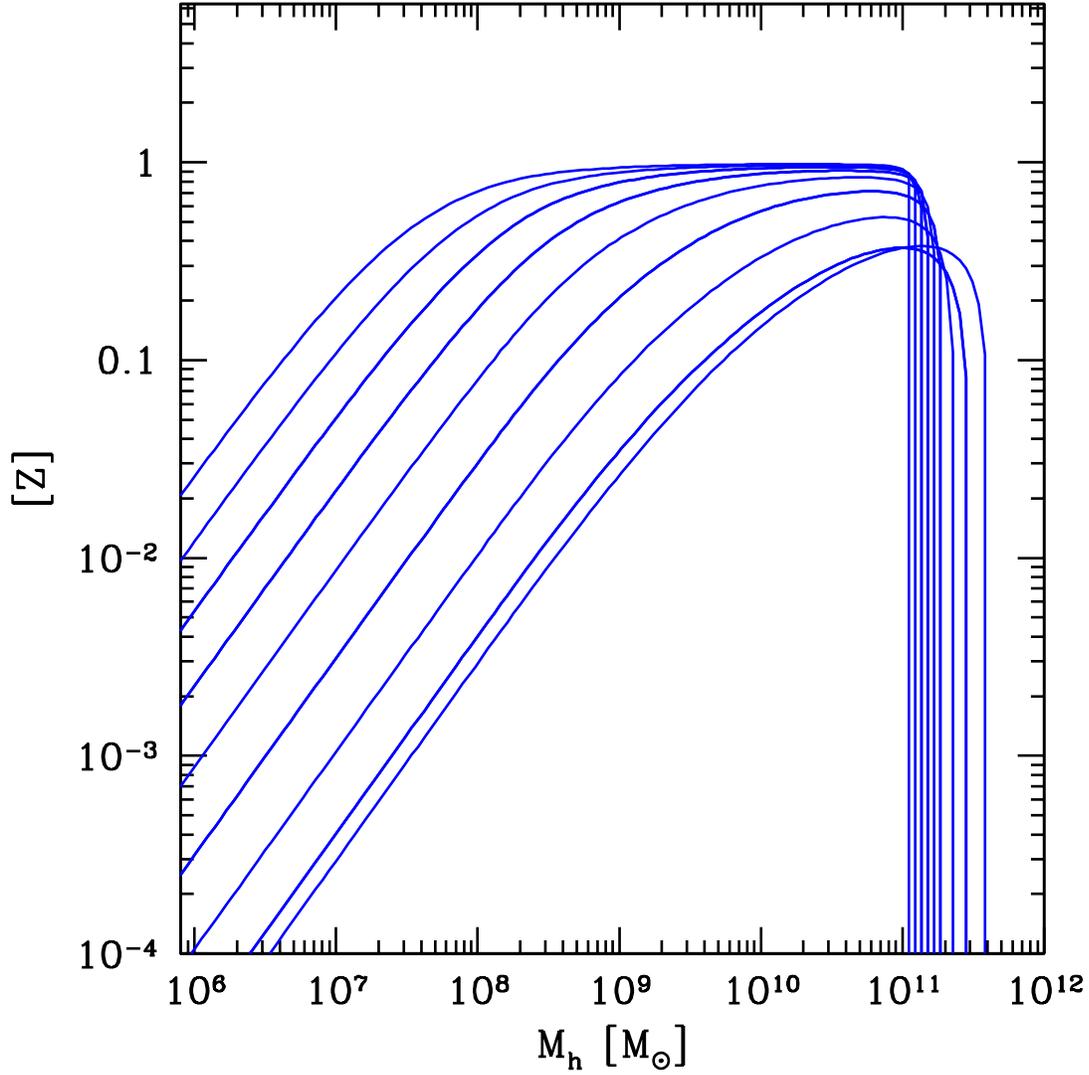,height=20cm}}
\caption{\label{fig4}\footnotesize{Metallicity of diffusive spheres as
a function of parent halo mass for redshifts $z=0, 0.5, 1, 1.5, 2, 2.5, 3, 3.5,
4$ from the bottom to the top curves, respectively. 
The curves shown are for the case where the initial metallicity of the hot gas in 
the superbubbles is solar.
}}
\end{figure}
\begin{figure}
\centerline{\psfig{figure=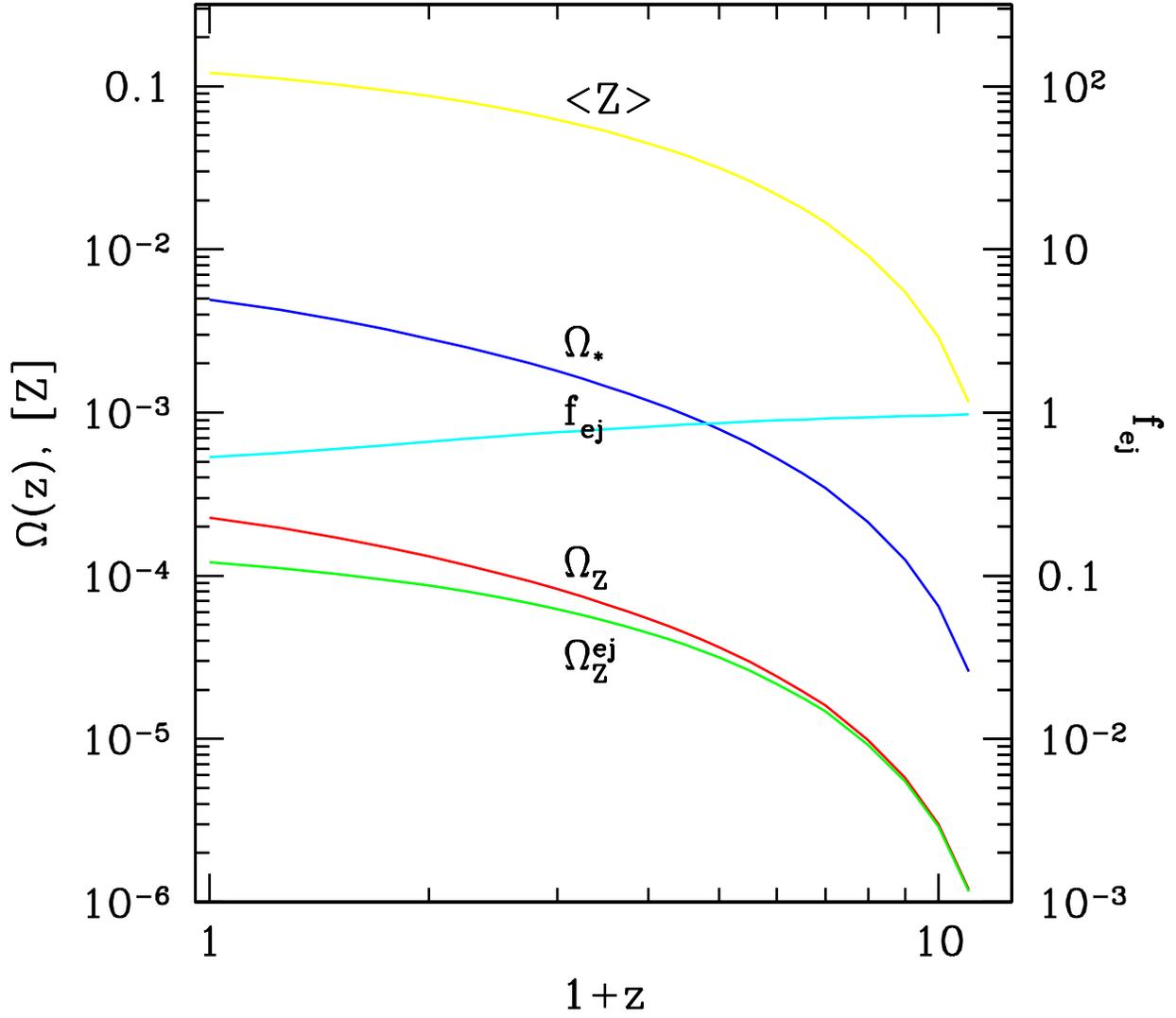,height=20cm}}
\caption{\label{fig5}\footnotesize{Redshift evolution of density parameter
for produced ($\Omega_Z$) and ejected ($\Omega_Z^{ej}$) metals, stars ($\Omega_\star$),
cosmic metal escape fraction ($f_{ej}$), and average IGM metallicity ($\langle Z
\rangle$) for the SCDM model.
}}
\end{figure}
\begin{figure}
\centerline{\psfig{figure=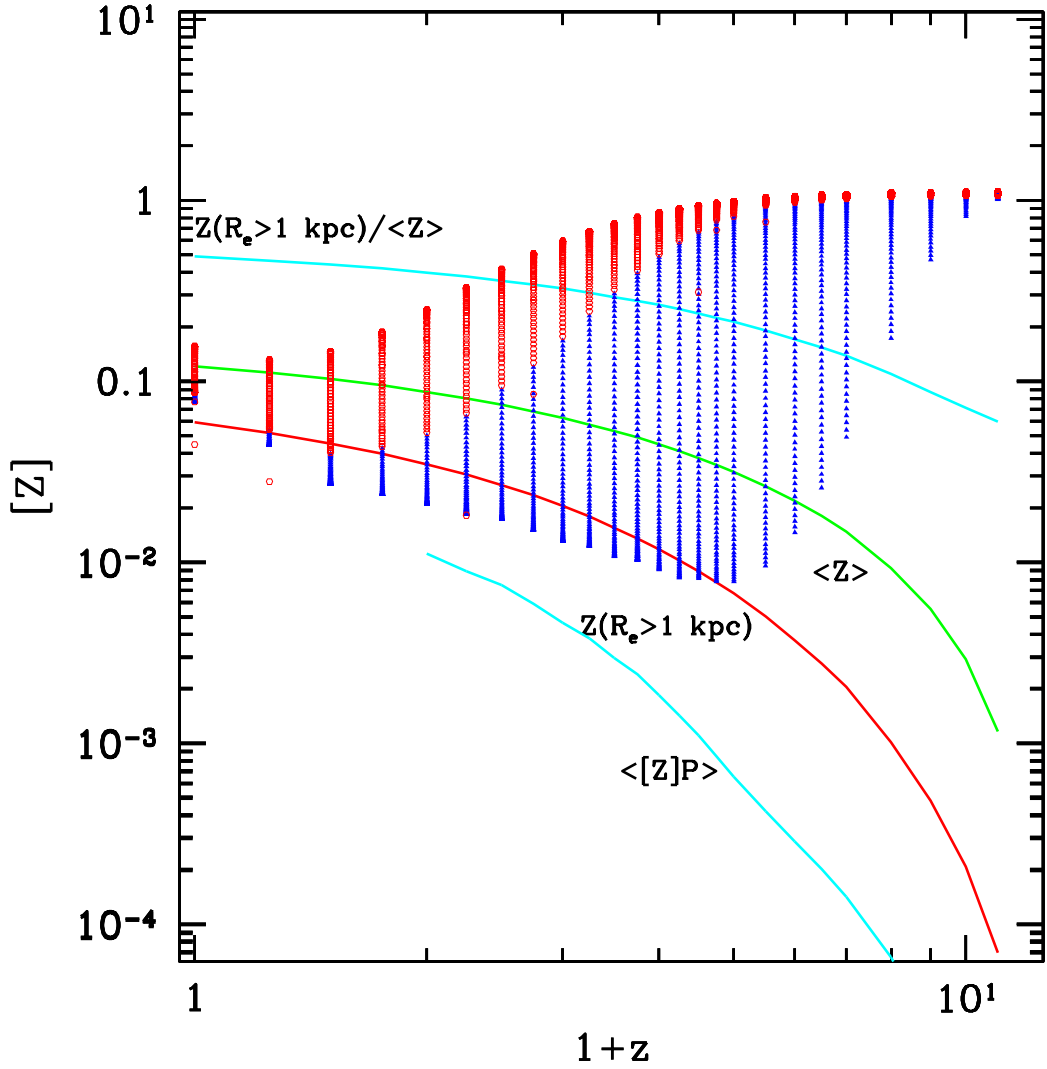,height=20cm}}
\caption{\label{fig6}\footnotesize{Redshift evolution of the IGM metallicity
distribution in the SCDM model. Also shown are the average IGM metallicity, $\langle Z \rangle$,
the contribution from objects with large metal-enriched spheres, $Z(R_e>1{\rm
~kpc})$, the ratio of the
two, and the covering factor-weighted metallicity, $\langle ZP(z)\rangle$.       
The initial value of the 
hot gas metallicity is $Z_\odot$ (standard case). 
}}
\end{figure}
\begin{figure}
\centerline{\psfig{figure=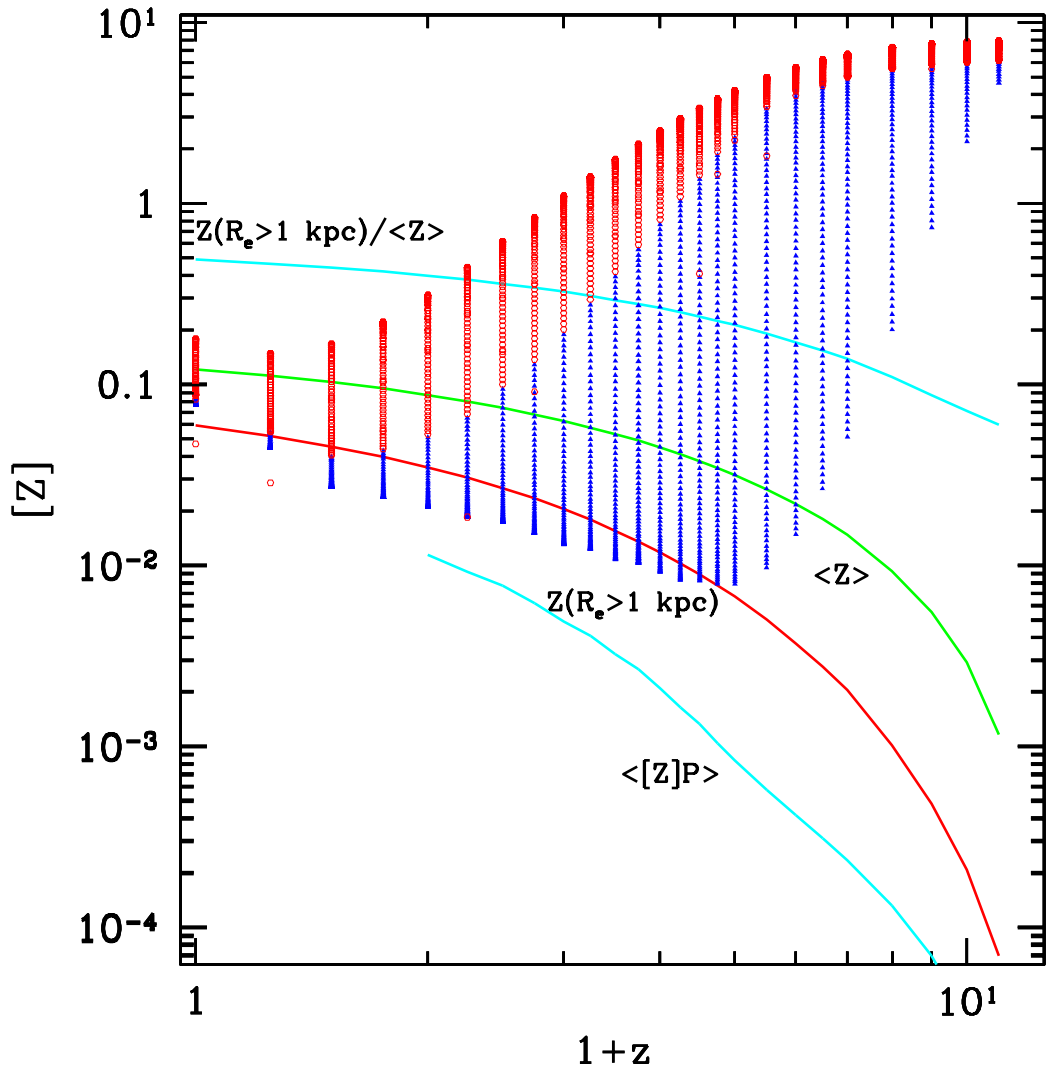,height=20cm}}
\caption{\label{fig7}\footnotesize{ Same as Fig. \ref{fig6}, but for  
initial value of the hot gas metallicity 8 $Z_\odot$.
}}
\end{figure}
\begin{figure}
\centerline{\psfig{figure=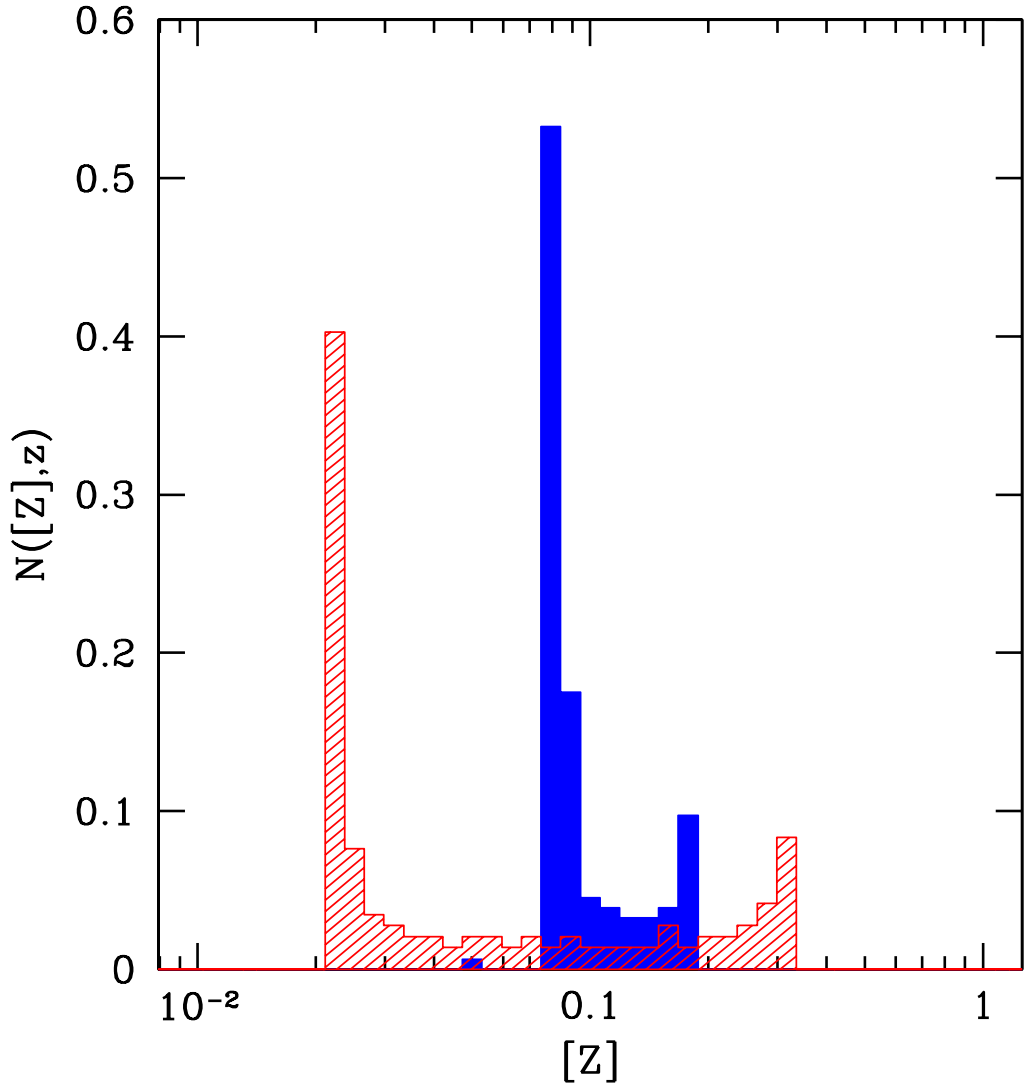,height=20cm}}
\caption{\label{fig8}\footnotesize{IGM metallicity distribution function for
 the SCDM model at
redshift $z=1$ (obliquely dashed) and $z=0$ (solid). Both histograms
are normalized to unity; the initial value of the hot gas metallicity 
is $Z_\odot$. 
}}
\end{figure}
\end{document}